\documentclass[twocolumn, 10pt, aps, pra, superscriptaddress]{revtex4-1}

\usepackage{amsmath,amssymb,mathtools,xcolor,bbold}
\usepackage{graphicx,bm}
\usepackage[T1]{fontenc}
\usepackage{hyperref}
\usepackage[utf8]{inputenc}
\usepackage{color}
\usepackage{braket}

\inputencoding{utf8}

\begin{document}

\title{Phase diagram of the dissipative quantum Ising model on a square lattice}

\author{Jiasen Jin}
\affiliation{School of Physics, Dalian University of Technology, 116024 Dalian, China}

\author{Alberto Biella}
\affiliation{Laboratoire Mat\'{e}riaux et Ph\'{e}nom\`{e}nes Quantiques, Universit\'{e} Paris Diderot, CNRS-UMR7162, 75013 Paris, France}

\author{Oscar Viyuela}
\affiliation{Department of Physics, Massachusetts Institute of Technology, Cambridge, MA 02139, USA}
\affiliation{Department of Physics, Harvard University, Cambridge, MA 02318, USA}

\author{Cristiano Ciuti}
\affiliation{Laboratoire Mat\'{e}riaux et Ph\'{e}nom\`{e}nes Quantiques, Universit\'{e} Paris Diderot, CNRS-UMR7162, 75013 Paris, France}
  
\author{Rosario Fazio}
\affiliation{ICTP, Strada Costiera 11, I-34151 Trieste, Italy}
\affiliation{NEST, Scuola Normale Superiore and Istituto Nanoscienze-CNR, I-56126 Pisa, Italy}

\author{Davide Rossini}
\affiliation{Dipartimento di Fisica, Universit\`a di Pisa and INFN, Largo Pontecorvo 3, I-56127 Pisa, Italy}

\date{\today}

\begin{abstract}
  The competition between interactions and dissipative processes in a quantum
  many-body system can drive phase transitions of different order.
  Exploiting a combination of cluster methods and quantum trajectories, we show how the systematic inclusion
  of (classical and quantum) nonlocal correlations at increasing distances is crucial to determine the structure
  of the phase diagram, as well as the nature of the transitions in strongly interacting spin systems.
  In practice, we focus on the paradigmatic dissipative quantum Ising model:
  in contrast to the non-dissipative case, its phase diagram is still a matter of debate in the literature.
  When dissipation acts along the interaction direction, we predict important quantitative modifications 
  of the position of the first-order transition boundary. 
  In the case of incoherent relaxation in the field direction, our approach confirms the presence
  of a second-order transition, while does not support the possible existence of multicritical points.
  Potentially, these results can be tested in up-to-date quantum simulators of Rydberg atoms.
\end{abstract}


\maketitle

\section{Introduction}

Quantum phase transitions are a cornerstone of modern statistical mechanics,
originating when the ground state of a many-body system changes either continuously
or abruptly, in virtue of a nonthermal control parameter~\cite{Sachdev_book}.
This paradigm substantially changes in out-of-equilibrium conditions,
where thermodynamic equilibrium is absent and all energy levels become relevant.
A way to witness such non-equilibrium phenomena is, e.g., by considering an isolated quantum system
and studying how it reacts to an abrupt change in one of its parameters~\cite{Polkovnikov_2011}. 
Depending on its spectral properties, the system can locally thermalize to some equilibrium ensemble
or get stuck in a more exotic many-body localized phase~\cite{Huse_2015}.
Alternatively, a system can be driven away from equilibrium by putting it in contact
with an external environment which is at odds with the Hamiltonian dynamics
(and thus does not induce thermalization).
In such case the dynamics is non-unitary and, after an initial transient time,
may end up in a (possibly mixed) steady state, losing track of the initial conditions.
Here we prove how the build up of classical and quantum correlations dramatically modifies the nature of phase transitions
in open systems. By employing a combination of state-of-the-art numerical approaches,
we explore how these nonequilibrium systems behave near criticality.
In particular, we concentrate on a prototypical quantum Ising spin-$1/2$ system coupled to different Markovian 
(memoryless) environments, whose essential properties can be captured
by a Liouvillian master equation in the Lindblad form~\cite{Petruccione_book, Rivas_book}.

The amazing possibilities offered by several experimental platforms,
as atomic and molecular optical systems~\cite{Muller_2012}, arrays of coupled QED cavities~\cite{Houck_2012, Fitzpatrick_2017},
or coupled optomechanical resonators~\cite{Ludwig_2013}, recently spurred a considerable theoretical interest
in the investigation of quantum matter under such framework, including the emergence of critical phenomena and collective behaviors.
In view of the complexity of the problem and the rarity of exactly solvable models~\cite{FossFeig_2017},
several analytical and numerical methods have been developed in order to deal with systems in two (or more)
spatial dimensions, where critical phenomena are most likely to occur (see for example
Refs.~\cite{Degenfeld_2014, Weimer_2015, Finazzi_2015, Jin_2016, Orus_2016, Sieberer_2016, Li_2016, Mascarenhas_2017, Casteels_2018, Traj_2018, Biella_2018, Nagy_2018, Rota_2018}).
However their general classification is still at its infancy~\cite{Minganti_2018}. 

In this paper we shed new light on the impact of correlations in dissipative quantum phase transitions,
motivated by the recent realization of a programmable
quantum spin model with tunable interactions~\cite{Bernien_2017, Keesling_2018}. While on typical experimental time scales
the dynamics can be safely approximated as unitary, it is possible to enhance the dissipation channel,
such to compete with the Hamiltonian dynamics, by coupling the Rydberg state to short-lived auxiliary energy levels.
Through extensive numerical calculations, we highlight how the systematic inclusion of classical and quantum correlations
at increasing distances is crucial to determine the structure of the phase diagram,
as well as the nature of the critical boundaries in strongly interacting spin systems.
In particular we exploit cluster approaches, which have been shown to provide quantitatively accurate results in the description
of the phase diagram and of critical phenomena in dissipative quantum lattice systems~\cite{Jin_2016, Biella_2018}.
For the sake of concreteness, we frame our analysis in the paradigmatic
transverse-field Ising model, with dissipation in the form of incoherent spin flips. 
This has been object of intense theoretical investigation~\cite{Lee_2011, Lee_2013, Marcuzzi_2014, Rota_2017, Roscher_2018},
in view of its direct experimental simulation with interacting Rydberg atoms~\cite{Carr_2013, Lienhard_2018, Leseleuc_2018, Barredo_2018}.
The resulting steady-state phase diagram in two dimensions raised a number of debated issues on the nature
of the various transitions~\cite{Weimer_2015, Orus_2016, Maghrebi_2016, Rose_2016, Overbeck_2017},
and constitutes the main focus of the present paper.

We provide evidence that, depending on the choice of the privileged axis for incoherent flips,
the system exhibits either first-order or continuous transitions.
When dissipation acts along the (internal) direction of spin-spin interaction, we show how the known mean-field (MF) 
bistability phenomena translate into first-order transitions or smooth crossovers according to the interaction strength. 
Our predictions quantitatively modify the phase diagram structure with respect to the one reported
in the literature~\cite{Weimer_2015}. 
In the case of incoherent relaxation in the field direction, our approach does not support the possible existence
of a multicritical point~\cite{Overbeck_2017}, unveiling how the emerging transition is always of second order. 
The effect of interactions is highlighted by characterizing the correlation length across
first- and second-order transitions.

\section{Model}

The spin system Hamiltonian under investigation, ruling the coherent part of the dynamics, is
\begin{equation}
  \hat{H} = \frac{V}{4} \sum_{\langle j,l \rangle} \hat{\sigma}^z_j \hat{\sigma}^z_l + \frac{g}{2} \sum_j \hat{\sigma}^x_j ,
\end{equation}
where $\hat{\bm \sigma}_j \equiv (\hat\sigma^x_j,\hat\sigma^y_j,\hat\sigma^z_j)$
denote the spin-$1/2$ Pauli matrices on site $j$ of a two-dimensional square lattice. 
The first term represents the nearest-neighbor interaction along $z$ of strength $V$, while the second term accounts for 
a local and uniform magnetic field along the transverse direction $x$.
We consider two different kinds of incoherent dissipative processes, acting
independently and locally on each spin: these tend to flip it down either along
the coupling ($z$), or along the field ($x$) direction.
The full master equation governing the evolution of the system's density matrix $\rho(t)$ is
\begin{equation}
  \partial_t \rho (t) 
  = -{\rm i} [\hat{H},\rho] + \gamma \sum_j \bigg( \hat{L}_j \rho \hat{L}_j^\dagger - \frac12 \{ \hat{L}^\dagger_j \hat{L}_j, \rho \} \bigg),
  \label{eq:Master}
\end{equation}
where the Lindblad jump operators on each site are all given by either
$\hat{L}^{(z)}_j \equiv \tfrac12(\hat\sigma^x_j - {\rm i} \hat\sigma^y_j)$
or $\hat{L}^{(x)}_j \equiv \tfrac12(\hat\sigma^z_j -{\rm i} \hat\sigma^y_j)$, respectively.
In what follows, we will be interested in the long-time limit properties of Eq.~\eqref{eq:Master},
which is guaranteed to reach a unique steady state (SS): $\rho_{\rm SS} = \lim_{t \to \infty} \rho(t)$~\cite{Schirmer_2010}.
Hereafter we adopt units of $\hbar = 1$.

\section{Results}

\subsection{Model I: Dissipative spin flips along the coupling direction}
\label{sec:modelI}

\begin{figure}[!t]
  \centering
  \includegraphics[width=1.\columnwidth]{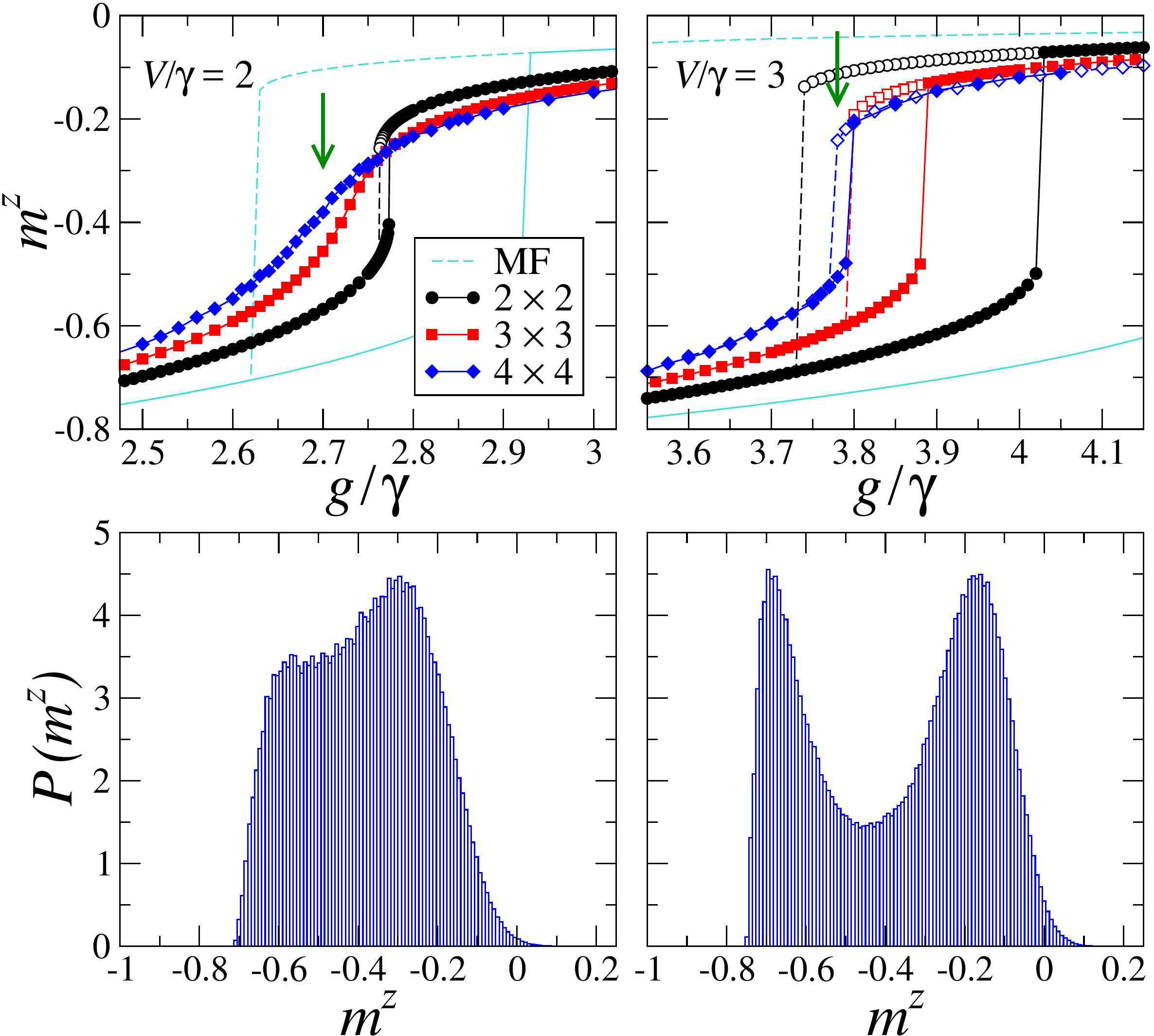}
  \caption{
    {\it Model I.} --- 
    Upper panels: average steady-state magnetization $m^z$ as a function of the transverse field $g/\gamma$,
    for two different values of the coupling, $V/\gamma = 2 < V_c/\gamma$ (left)
    and $V/\gamma =3 \gtrsim V_c/\gamma$ (right).
    Light blue lines indicate the two branches of the MF solution.
    Symbols are results of CMF simulations for clusters of various size, as indicated in the legend.
    The upper/lower branch is denoted by empty/filled symbols.
    Lower panels: histogram of $m^z$ sampled in time by a single quantum trajectory
    in a cluster with $\ell = 4$, for $V/\gamma=2, \, g/\gamma=2.7$ (left)
    and for $V/\gamma=3, \, g/\gamma=3.78$ (right).}
  \label{fig:Fig1}
\end{figure}

Let us start by taking $z$-oriented dissipative spin flips ({\it Model I}).
A simple MF approach, based on a product ansatz $\rho(t) = \otimes_j \rho_j(t)$,
would decouple the Hamiltonian term
$q^{-1} \sum_{\langle j,l \rangle} \hat\sigma^z_j \hat\sigma^z_l \to m^z \sum_j \hat\sigma^z_j$,
where $m^z = {\rm Tr} \big[ \hat \sigma^z_j \, \rho_{\rm SS} \big] \equiv \langle \hat\sigma^z_j \rangle$
is the local magnetization and $q$ is the coordination number of the lattice,
thus splitting the many-body Liouvillian into a sum of single-spin operators which can then be
treated straightforwardly in a self-consistent way.
Under this approximation, according to the choice of the system parameters, in some cases the solution
is not unique and a bistable behavior (in which $\rho_{\rm SS}$ depends on the initial condition) emerges.
This clearly appears in the observable $m^z$, as displayed by the light blue lines
in the upper panels of Fig.~\ref{fig:Fig1}, where the two branches correspond to solutions
obtained when sweeping from larger to smaller values of $g$ (dashed), or conversely (continuous lines).

A more careful analysis however admits an exact treatment of short-range correlations that may establish
within a cluster ${\cal C}$ of spins, while the MF is applied at the boundary of such cluster~\cite{Tomadin_2010, Jin_2016}.
We performed such a cluster mean-field (CMF) study in a lattice of dimension $\ell \times \ell$,
for clusters up to $\ell = 4$~\cite{footnote1}.
The resulting magnetization is shown in the same panels, with different symbols corresponding to various cluster sizes.
It appears quite neatly that the bistability region progressively shrinks when
increasing $\ell$, and eventually tends to disappear.
Specifically, for $\ell=4$, we can identify a threshold value of the coupling strength $V_c^{(\ell=4)} /\gamma \approx 3$,
separating a region where the magnetization asymptotically exhibits a continuous behavior as $g/\gamma$ is increased
(top left panel, for $V/\gamma = 2$), from another one where a discontinuity in $m^z$ spotlights the presence
of a first-order transition (top right panel, for $V/\gamma = 3$, where the putative transition is located at $g_c/\gamma \approx 3.78 \pm 0.05$).
However in the latter case, for the largest available cluster size, we are still observing resilience of the
system to bistability at long times, in a narrow region $3.78 \lesssim g/\gamma \lesssim 3.8$~\cite{footnote2}.
A rough finite-size scaling of data for $\ell \leq 4$ suggests the onset of a critical point
at a finite value ($V_c/\gamma \approx 4.05, \, g_c/\gamma \approx 4.88$), obtained by extrapolating
to the thermodynamic limit (see App.~\ref{app:modelI_PD}).
We shall emphasize that
the only other prediction available in the literature has been obtained using a variational approach~\cite{Weimer_2015}
that locates the critical point at $V_c/\gamma=1.4$, for which the transition is observed at $g_c/\gamma=2.28$.
Such estimate qualitatively agrees with our bare MF data, which slightly underestimate the location of $V_c$,
while completely washing out correlations between the various sites.
The full CMF phase diagram drawn in the $V$-$g$ plane is presented in App.~\ref{app:modelI_PD}.

Evidence for the change of behavior when crossing $V_c$ is also witnessed by analyzing the quantum jumps
that appear when monitoring the time evolution of a single stochastic trajectory~\cite{Vicentini_2018, Minganti_2018}.
Once the values of the couplings are fixed, each quantum trajectory explores states with different magnetization during its dynamics.
By sampling the outcomes, one obtains an histogram representing the probability to measure
a given value of $m^z$ in typical quantum optical experiments.
If $V>V_c$, the system jumps abruptly from the low- to the high-density phase
as $g$ is increased (see App.~\ref{app:bistab} for details).
By performing such a measure at the critical point $g_c$, the probability distribution turns out to be bimodal,
since the trajectory mainly jumps between the two phases (lower right panel of Fig.~\ref{fig:Fig1}).
In particular, it is the symmetric sum on the probability distributions one would obtain
in the two phases for $g<g_c$ and $g>g_c$, respectively.
This reflects the fact that, at criticality, the density matrix must be the equiprobable mixture of the two phases~\cite{Minganti_2018}. 
For $V<V_c$ the system exhibits a smooth crossover between the two phases.
By applying the same protocol for $g/\gamma=2.7$ (the value for which $\partial_g m^z(g)$ is larger),
the bimodal character of the distribution is smeared out, thus signaling the disappearance
of the critical behavior (lower left panel of Fig.~\ref{fig:Fig1}).

\begin{figure}[!t]
  \centering
  \includegraphics[width=1.\columnwidth]{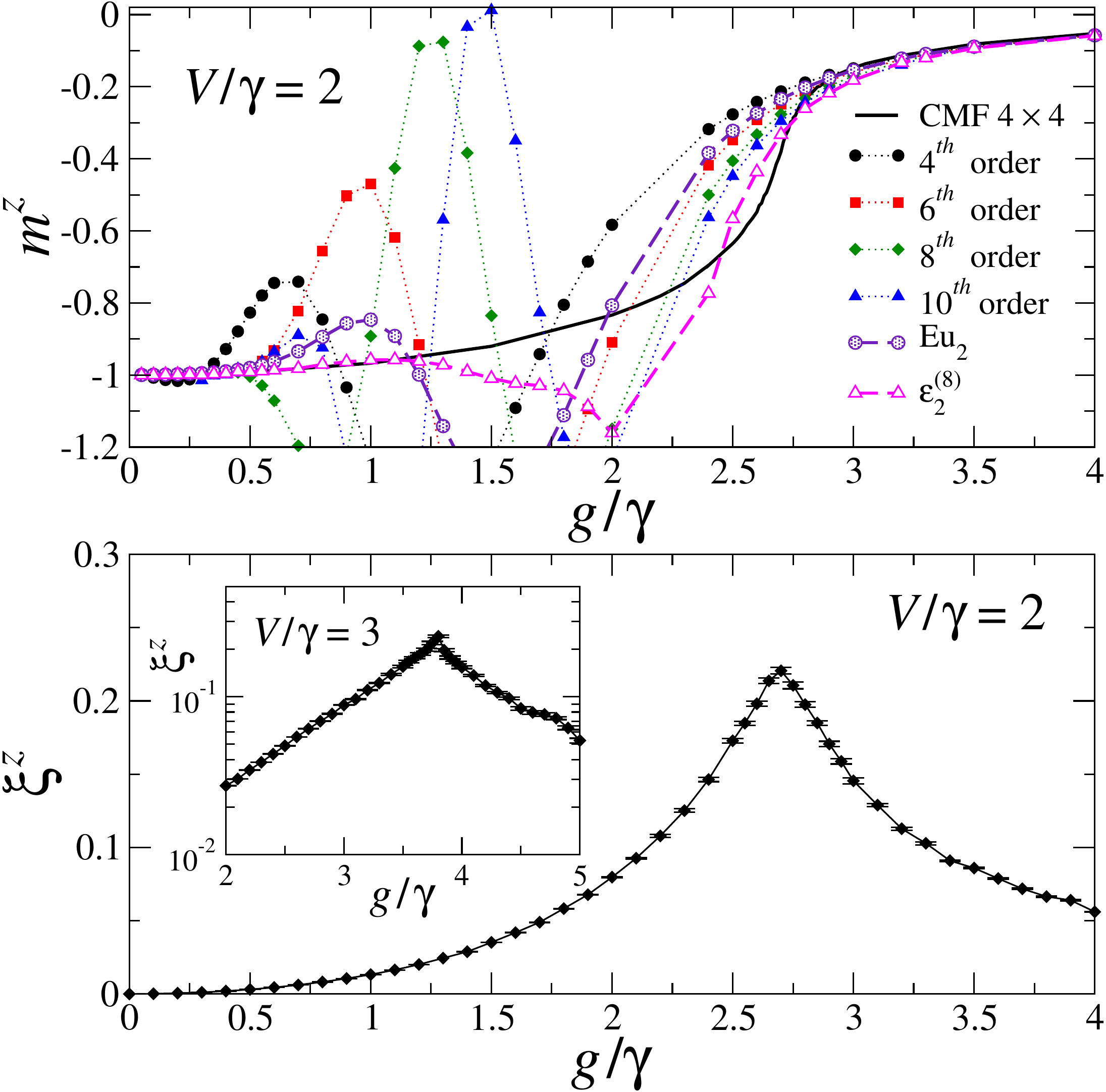}
  \caption{{\it Model I.} --- Upper panel: magnetization as a function of $g/\gamma$ for $V/\gamma=2$,
    evaluated with a NLCE up to $10$-th order and after using resummation techniques~\cite{footnote4}.
    Lower panel: correlation length $\xi^z$ in a $4 \times 4$ cluster
    for $V/\gamma = 2$ (main frame, linear scale) and $V/\gamma =3$ (inset, logarithmic scale).}
  \label{fig:Fig2}
\end{figure}

To get further insight about the impact of correlations on this physics, we employed a numerical
linked cluster expansion (NLCE)~\cite{Rigol_2006, Biella_2018}.
A remarkable advantage of NLCE over other strategies is that it enables a direct access to the thermodynamic limit of an infinite number of spins,
up to order $R$ in the cluster size, by only counting cluster contributions of sizes equal or smaller than $R$.
Importantly, contrary to other perturbative expansions, the NLCE is not based on a perturbative parameter
and the convergence of the series is controlled by the typical length scale of correlations~\cite{Rigol_2006, Biella_2018}.
In the top panel of Fig.~\ref{fig:Fig2}, we perform a NLCE for the observable $m^z(g)$ for $V/\gamma=2$ (crossover region).
As is clear by looking at the expansion truncated at different orders (see the legend), the series does not converge
in the crossover region, even when resummation techniques to speed up the convergence are employed~\cite{footnote4}.
This is due to a dramatic build up of correlations in the crossover region, even in the non critical case.

\begin{figure*}[!t]
  \centering
  \includegraphics[width=0.68\columnwidth]{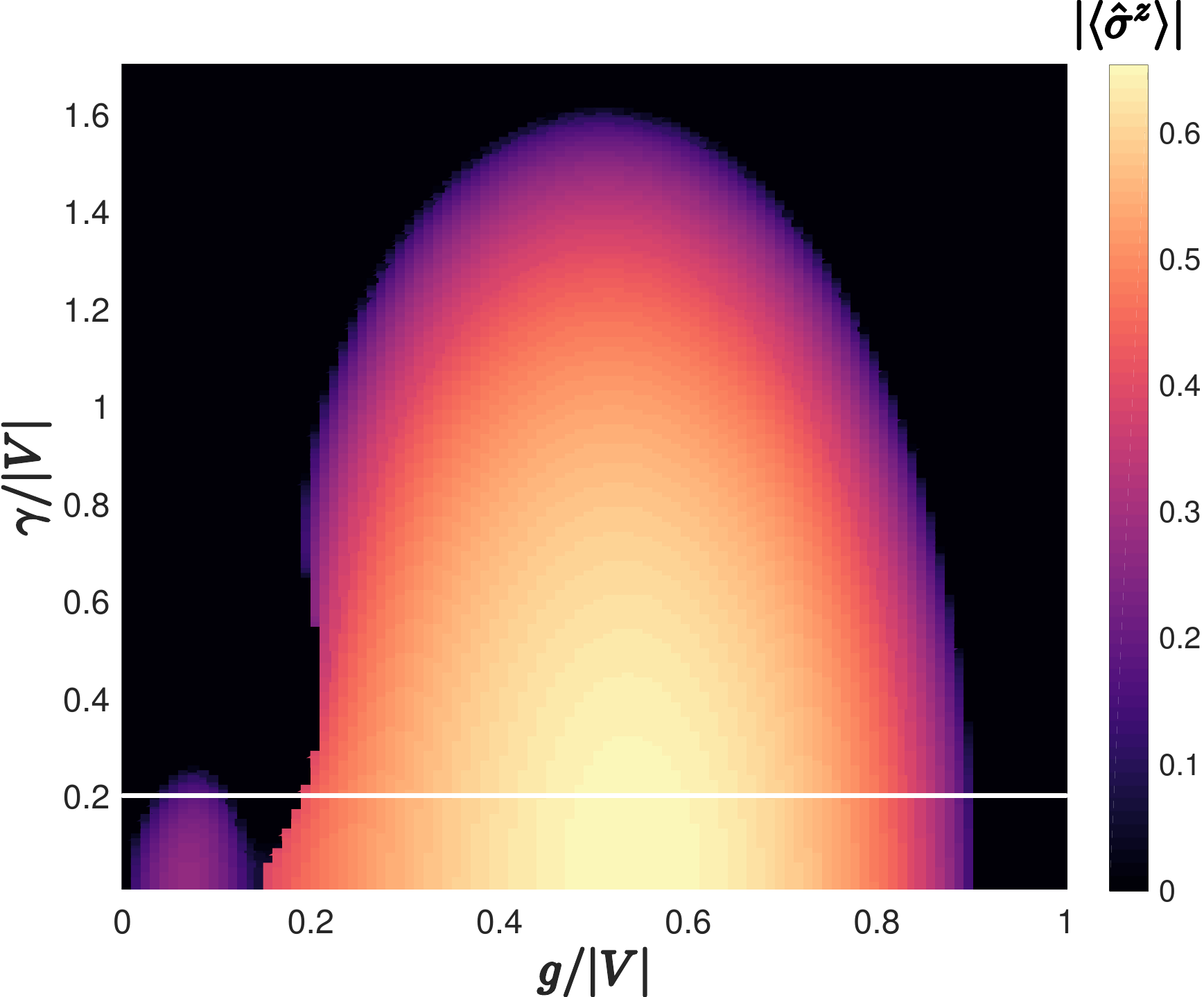}
  \includegraphics[width=0.68\columnwidth]{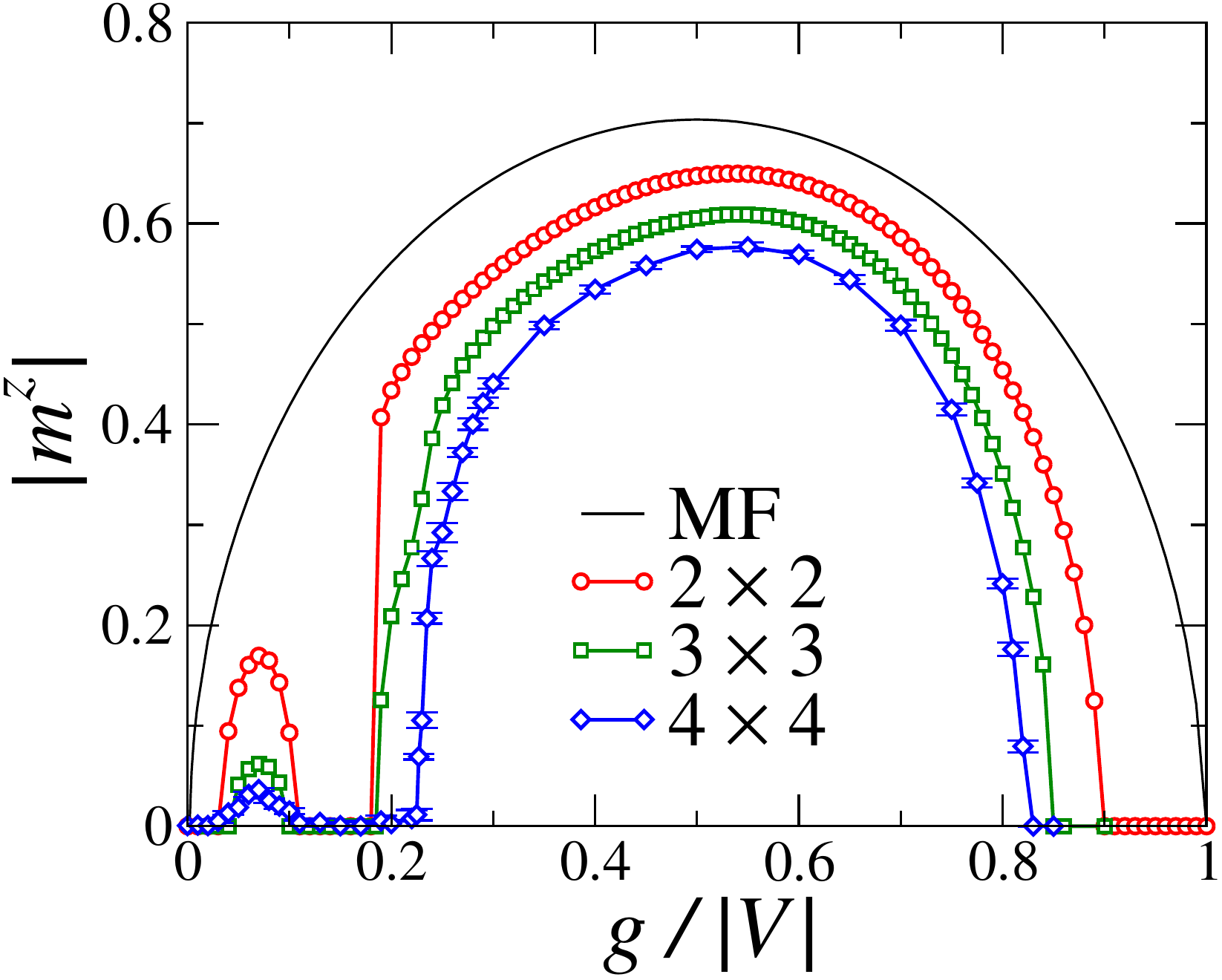}
  \includegraphics[width=0.68\columnwidth]{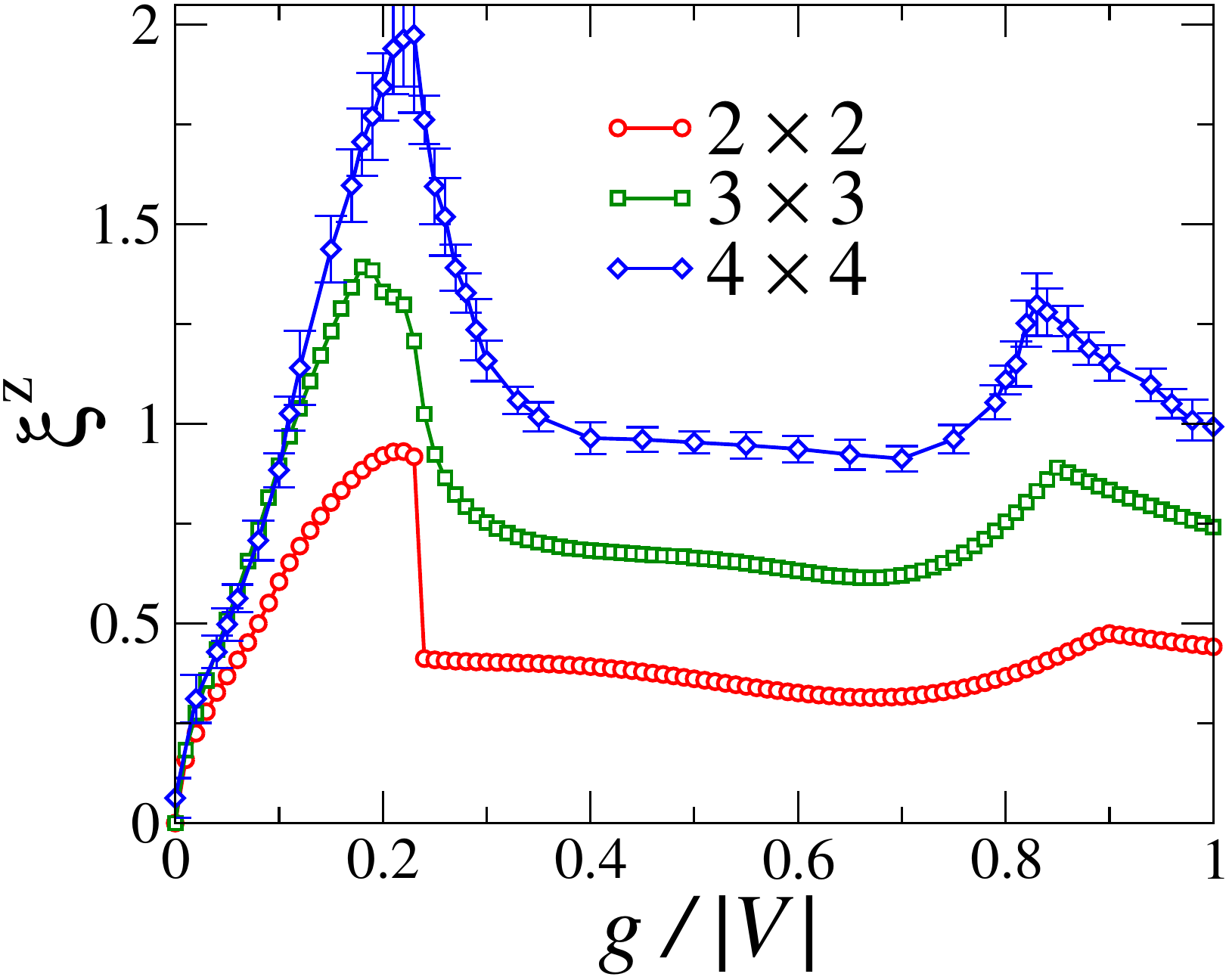}
  \caption{
    {\it Model II.} --- 
    Left panel: steady-state phase diagram in the $\gamma/|V|$-$g/|V|$ plane, obtained with a $\ell =2$ CMF approach,
    witnessed by the average magnetization $|m^z| = |\langle \sigma^z \rangle|$. 
    The horizontal white line marks the cut at $\gamma/|V|=0.2$, which has been scrutinized in the central and right panels,
    where we considered CMF with clusters of different sizes (see the legend).
    Namely, they display the behavior of the order parameter $|m^z|$ (central panel)
    and of the correlation length $\xi^z$ (right panel, see Eq.~\ref{corrlength}).}
  \label{fig:model2}
\end{figure*}

We further analyzed this mechanism by studying the behavior, in the transition region,
of the correlation length~\cite{Campostrini_2014}
\begin{equation}
  \label{corrlength}
  \big( \xi^z \big)^2 = \frac{1}{N} \sum_{{\bf r}, {\bf r'}} |{\bf r} - {\bf r'}|^2 g({\bf r},{\bf r'}),
\end{equation}
where $g({\bf r},{\bf r'}) = \langle \hat \sigma^z_{\bf r} \, \hat \sigma^z_{\bf r'} \rangle
- \langle \hat \sigma^z_{\bf r} \rangle \langle \hat \sigma^z_{\bf r'} \rangle$
is the connected part of the two-point correlation function (specified by the coordinates ${\bf r},{\bf r'}$)
along the coupling direction, and $N$ is the number of sites.
While the above quantity refers to a correlation length calculated within a given (small) cluster,
and thus cannot be directly related to the convergence of NLCE in the thermodynamic limit,
it provides an intuition of the key role played by correlations close to critical points (or their precursors).
The results of calculations on a cluster of size $\ell = 4$ are displayed in the lower panel, for $V/\gamma=2$ (main frame)
and $V/\gamma=3$ (inset), where it is shown that $\xi^z$
undergoes a sudden increase in proximity of the transition point.
This behavior also occurs in the noncritical case $V<V_c$, highlighting the importance
of the exact treatment of short-range interactions even in a system that does not display a critical behavior. 
However we shall stress that, in the critical case, $\xi^z$ does not necessarily diverge in the thermodynamic limit $N\to\infty$
(as is the case for a second-order transition) and, for the clusters we were able to reach,
it takes relatively small values ($\approx 10^{-1}$).
This fact may hinder its experimental detection, even if such quantity can become arbitrarily large,
thus indicating the strongly correlated nature of the steady state.

\subsection{Model II: Dissipative spin flips along\\ the field direction}
\label{sec:modelII}

We now switch to $x$-oriented dissipative spin flips ({\it Model II}).
When incoherent processes take place along the field direction (i.e., orthogonal to the spin-spin coupling),
the physics of the model changes qualitatively.
This scheme has been first devised and studied in one dimension, where the onset of interesting steady-state
correlations and edge effects have been witnessed~\cite{Bardyn_2012, Joshi_2013}.
In higher dimensions and at MF level, the system undergoes a continuous transition from a disordered paramagnetic phase
($\braket{\hat \sigma^z}=0$) to a ferromagnetic state ($\braket{\hat \sigma^z}\neq0$) in which
the $\mathbb{Z}_2$-symmetry $(\hat \sigma^y, \hat \sigma^z) \to -(\hat \sigma^y, \hat \sigma^z)$ is spontaneously broken.
A more refined treatment based on the Keldysh formalism predicts the transition to be of first order
(with symmetry breaking) at sufficiently strong dissipation~\cite{Maghrebi_2016}.
According to a subsequent study with a variational ansatz~\cite{Overbeck_2017},
the transition can be either continuous or first order, depending of the dissipation rate $\gamma$;
the continuous and first-order transition lines meet at a tricritical point. 
Here we show that systematically including the effect of correlations at an increasing distance
leads to some modifications of the phase diagram, where only continuous transitions are present,
thus excluding the possibility of a multicritical behavior. We point out that an exact treatment of
short-range correlations seems to be crucial to the description of dissipative spin systems,
since they have been proven to be able to substantially change the phase-diagram structure~\cite{Jin_2016},
in accordance with alternative non-perturbative approaches based on tensor-network simulations~\cite{Orus_2016}.

To this aim, we study the two-dimensional phase diagram in the $\gamma/|V|$-$g/|V|$ plane, for $V<0$.
In the left panel of Fig.~\ref{fig:model2}, we show the absolute value of the average magnetization $|m^z|$
as obtained with a CMF approach for $\ell=2$.
The results agree with those in Ref.~\cite{Overbeck_2017}: for $\gamma/|V| \lesssim0.5$ the transition is
first-order, otherwise it is continuous.
In the central panel, we study the effect on $|m^z|$ of the exact inclusion of correlations at increasing distance,
by considering clusters of size $\ell \leq 4$.
A larger-cluster ansatz progressively smoothens the first-order jump, thus leading to a continuous transition.
The lobe appearing at small $g$ and small $\gamma$ is quickly suppressed as $\ell$ is increased,
and represents an artifact of the CMF ansatz, as also witnessed by a linear stability analysis (see App.~\ref{app:modelII}).
The figure displays numerical results obtained for the cut at $\gamma/|V|=0.2$,
but the same conclusions apply for the whole range $\gamma/|V|\lesssim0.5$ (not shown). 
In the right panel of Fig.~\ref{fig:model2}, we show the behavior of the correlation length in Eq.~\eqref{corrlength}
along the same cut ($\gamma/|V|=0.2$), for different cluster sizes (namely $\ell = 2, 3, 4$).
The emergence of critical points is again witnessed by an abrupt increasing of $\xi^z$.  
In contrast to the case of the first-order transition studied in {\it Model I} (Fig.~\ref{fig:Fig2}),
here the correlation length at criticality is about two orders of magnitude larger
and the peak is more likely to be experimentally detected.
Indeed, in the case of continuous transitions with symmetry breaking, a divergence of $\xi^z$
in the thermodynamic limit is expected to occur.

\section{Conclusions}

We proved how the emergence of classical and quantum correlations
dramatically modifies the nature of dissipative phase transitions in strongly interacting spin systems. 
Applying a combination of cluster methods and quantum trajectories on a testbed spin-$1/2$ quantum Ising model
with incoherent spin flips, we demonstrated two key results. 
First, quantum phase bistability evolves into a crossover or a purely first-order phase transition,
if short-range interactions are properly accounted for. This allowed us to precisely locate
the position of the critical boundary, which has been a matter of debate in the literature.
Second, certain previously thought first-order transitions are indeed second-order when employing an exact treatment of interactions.
This points toward excluding the presence of a multicritical behavior originated by the dissipative dynamics in the quantum Ising model. 
Our results also contribute to a full and comprehensive characterization of the role of correlation functions
close to dissipative critical points, which represent one of the main challenges in the field of open many-body systems.

Unveiling the effect of disorder as well as disentangling the contribution of classical
and quantum correlations at phase-transition points are intriguing future directions,
that can be tackled within this framework. 
Experimentally, the fast development of quantum simulators using Rydberg atoms~\cite{Carr_2013, Bernien_2017, Lienhard_2018, Keesling_2018} 
stands as an exciting opportunity to test these predictions in the lab.

\acknowledgments

We thank A. Keesling, M. Lukin, F. Minganti, E. Vicari, and F. Vicentini for fruitful discussions.
We are grateful to J. Keeling for a critical reading of the manuscript.
We acknowledge the CINECA award under the ISCRA initiative, for the availability of high performance computing resources and support.
AB and CC acknowledge support from ERC (via Consolidator Grant CORPHO No.~616233).
OV thanks Fundacion Ramon Areces and RCC Harvard.
JJ acknowledges supports from the National Natural Science Foundation of China via No.~11605022, No.~11775040 and No.~11747317.

\appendix

\section{Phase diagram of Model I}
\label{app:modelI_PD}

As already pointed out in Ref.~\cite{Weimer_2015}, the master equation~\eqref{eq:Master}
with $\hat L_j = \tfrac12(\hat\sigma^x_j - {\rm i} \hat\sigma^y_j)$
does not exhibit any spin symmetry, as the $\mathbb{Z}_2$-symmetry
of the Hamiltonian $(\hat \sigma^y, \hat \sigma^z) \to -(\hat \sigma^y, \hat \sigma^z)$
is broken by the dissipative terms.
It is however possible to have a transition where an emergent symmetry arises,
being characterized by a sudden enhancement of the polarization
$m^z = {\rm Tr} [ \hat \sigma^z_j \, \rho_{\rm SS}]$ along the same direction ($z$)
of the dissipative spin flips. Indeed it may happen that, if the interaction $V$ exceeds
a given critical value $V_c$, when increasing $g$ the system undergoes a first-order (discontinuous)
transition from a low-polarized phase [$g < g_1(V)$] to a high-polarized phase [$g > g_2(V)$]
($g_1(V)$ and $g_2(V)$ set the the left and the right boundary of the bistable region for $V>V_c$).
This picture is qualitatively confirmed by our numerical cluster mean-field (CMF) simulations.

\begin{figure}[!t]
  \centering
  \includegraphics[width=0.95\columnwidth]{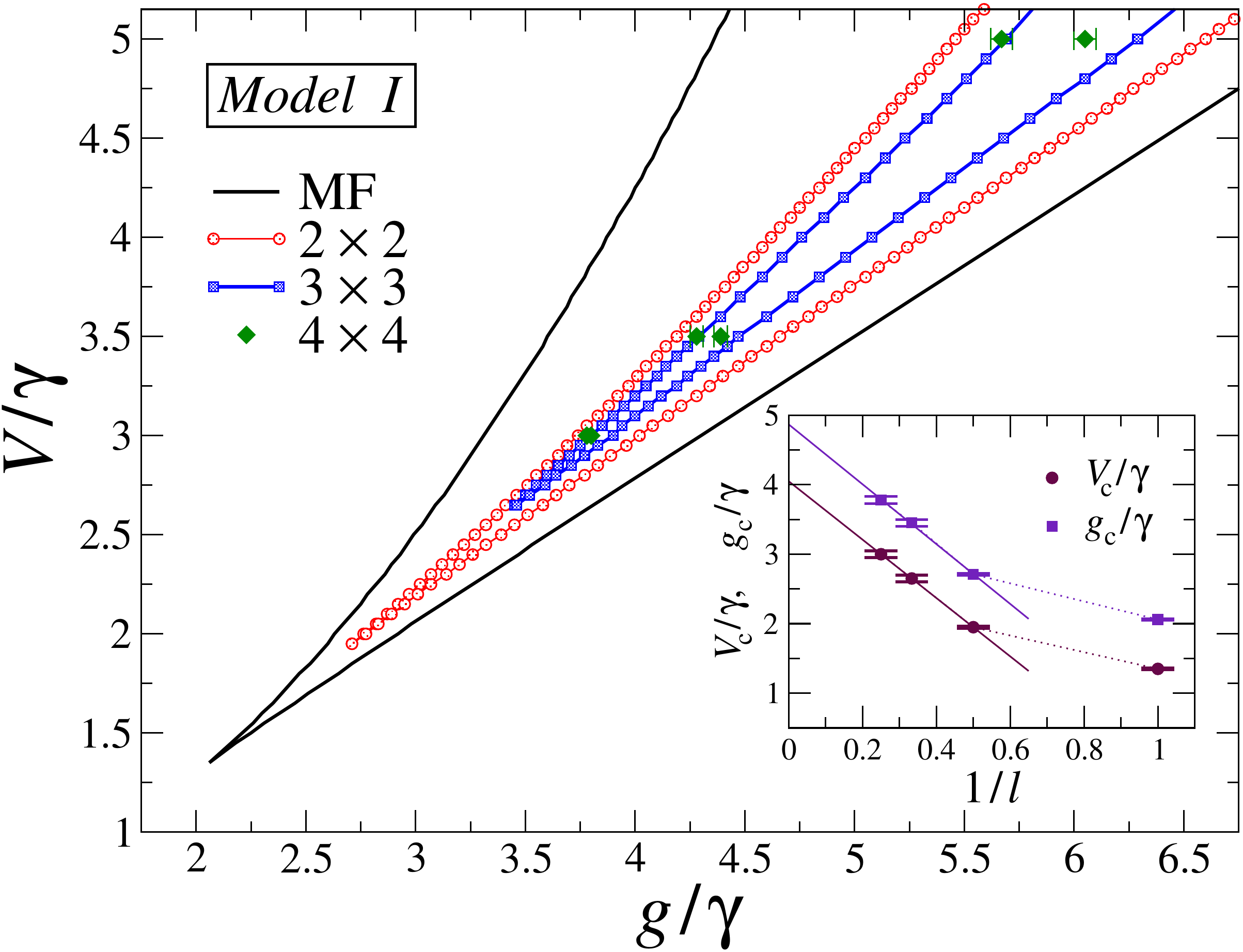}
  \caption{{\it Model I.} --- Steady-state phase diagram in the $V/\gamma$-$g/\gamma$ plane,
    as obtained by CMF simulations for different cluster sizes (see the legend: 
    various symbols and colors stand for different values of $\ell$).
    For a fixed value of $V/\gamma$ above a given threshold,
    lines of the same color separate two different phases with small magnetization
    (for $g<g_1$) and with large magnetization (for $g>g_2$).
    In between the two phases (for $g_1 < g < g_2$), a bistability region appears.
    The continuous black lines denote the single-site MF prediction.
    Inset: finite-size scaling analysis of the tip position with the cluster size.
    Data for $V_c^{(\ell)}/\gamma$ and $g_c^{(\ell)}/\gamma$ are plotted against $1/\ell$.
    Continuous lines are linear fits of the numerical data (symbols) for $\ell \geq 2$,
    which extrapolate, in the thermodynamic limit $\ell \to \infty$,
    to $V_c/\gamma \approx 4.05$ and $g_c/\gamma \approx 4.88$.}
  \label{fig:PhaseDiag}
\end{figure}

The resulting steady-state phase diagram of {\it Model I}, in the $V$-$g$ plane,
is drawn in Fig.~\ref{fig:PhaseDiag}.
This has been obtained using a CMF approach on clusters of size $\ell \times \ell$,
with different values $\ell = 2, \, 3, \, 4$.
We observe that, for $V/\gamma$ below a given threshold, the system does not exhibit any sharp
transition, while above that threshold a transition between low- and high-polarized phase occurs.
In between the two phases, the underlying MF approximation predicts a bistable region,
whose size shrinks as the cluster dimension $\ell$ is increased.
Bistability is characterized by the fact that the asymptotic steady-state solution,
and consequently the magnetization, is not unique and depends on the initial condition.
We ascribe the onset of bistability to an artifact of the MF decoupling, which should disappear
in the large-volume limit $\ell \to \infty$, where the low/high-polarized transition
truly becomes of first order (i.e., for $V>V_c^{(\ell)}$, whose location depends
on the size $\ell$, the magnetization $m^z$ exhibits a sudden jump
at a critical value $g_c^{(\ell)}$).
Notice that, with increasing $\ell$, the position of $V>V_c^{(\ell)}$ also shifts to larger values.

In the inset of Fig.~\ref{fig:PhaseDiag}, we report a finite-size scaling analysis of such position
with the cluster size $\ell$, where we observe a rough linear dependence of $V>V_c^{(\ell)}$
and of $g_c^{(\ell)}$ with $1/\ell$. Specifically, by linearly fitting our numerical
data for $\ell \geq 2$, we are able to predict the onset of a critical point for $\ell \to \infty$
at a finite position in the $V-\gamma$ plane (i.e., the tip in the bottom left side
of each data set, which progressively shifts to larger values of $V$ and $\gamma$).
This can be approximately located at
$V_c/\gamma \approx 4.05$, for which the putative first-order transition is found
at $g_c/\gamma = g_1/\gamma = g_2/\gamma \approx 4.88$.

\subsection{Emergence of bistability close to the transition}
\label{app:bistab}

Let us now have a closer look at the temporal behavior of the average magnetization
$m^z(t) = {\rm Tr}[\hat \sigma^z_j \, \rho(t)]$,
in proximity of the bistability region (as it appears in our CMF simulations).
The bistability phenomenon emerges in the fact that the steady state reached by time evolving
the master equation~\eqref{eq:Master} in a self-consistent way does depend on the initial conditions.

\begin{figure}[!t]
  \centering
  \includegraphics[width=0.95\columnwidth]{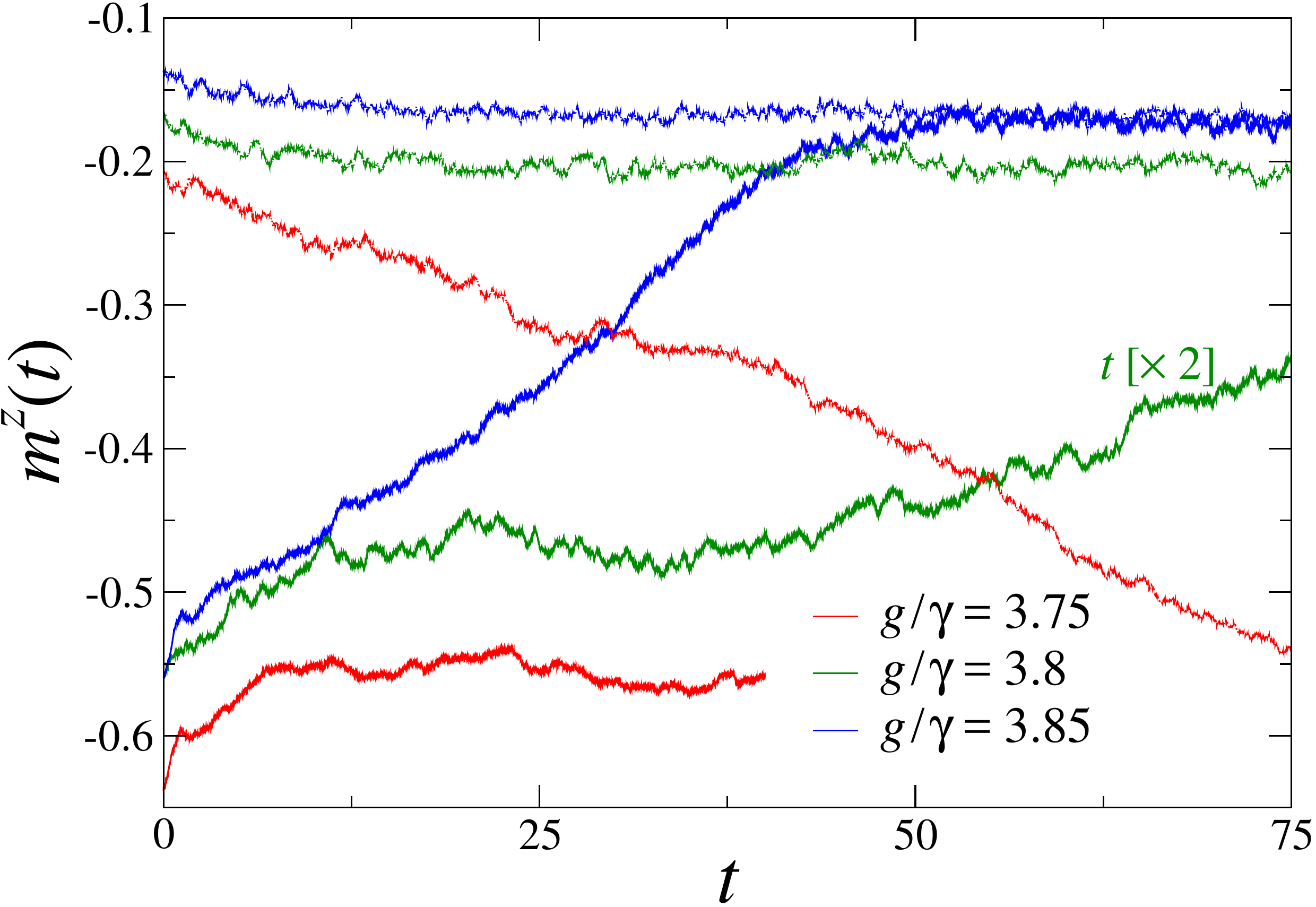}
  \caption{{\it Model I.} --- Time evolution of the average magnetization $m^z(t)$, at fixed $V/\gamma=3$.
    The data sets of various colors (see the legend) correspond to different values of $g/\gamma$, 
    which have been calculated by means of a CMF analysis with $\ell = 4$.
    Curves going up (continuous lines) have been obtained by integrating the master equation~\eqref{eq:Master}
    for a single cluster, starting from an initial condition corresponding to a value $g/\gamma = 3.7$.
    Curves going down (dotted lines) are for initial conditions corresponding to a value $g/\gamma = 3.9$.
    Note that the running time of the curve going up for $g/\gamma=3.8$ has to be multiplied by a factor $2$.}
  \label{fig:Conv_t}
\end{figure}

In practice we observe the following behavior, close to the bistable region.
Starting from the steady state reached for $g<g_1$ and searching the new steady-state CMF solution 
for increasing values of $g$, the needed convergence times dramatically increase for $g_1<g<g_2$.
In particular, we notice a strong tendency to stay in the low-magnetized configuration,
while the highly-magnetized phase is eventually reached only at $g \gtrsim g_2$ after waiting a long time.
An analogous phenomenology occurs in the same region, when starting from the steady state
reached for $g>g_2$ and searching the new solution for decreasing values of $g$.
This time the system tends to remain in the highly-magnetized configuration,
while the low-magnetized phase is only reached at $g \lesssim g_1$, after a long time.
Figure~\ref{fig:Conv_t} pinpoints this trend.

We point out that, usually, the convergence time of the simulations increases in proximity of the critical point;
this fact is related to the reduction of the Liouvillian gap, which eventually determines the longest time scale 
that is needed for approaching the steady state~\cite{Macieszczak_2016, Rose_2016}.

\subsection{Bimodality of single quantum trajectories}
\label{app:bimod}

\begin{figure}[!t]
  \centering
  \includegraphics[width=0.95\columnwidth]{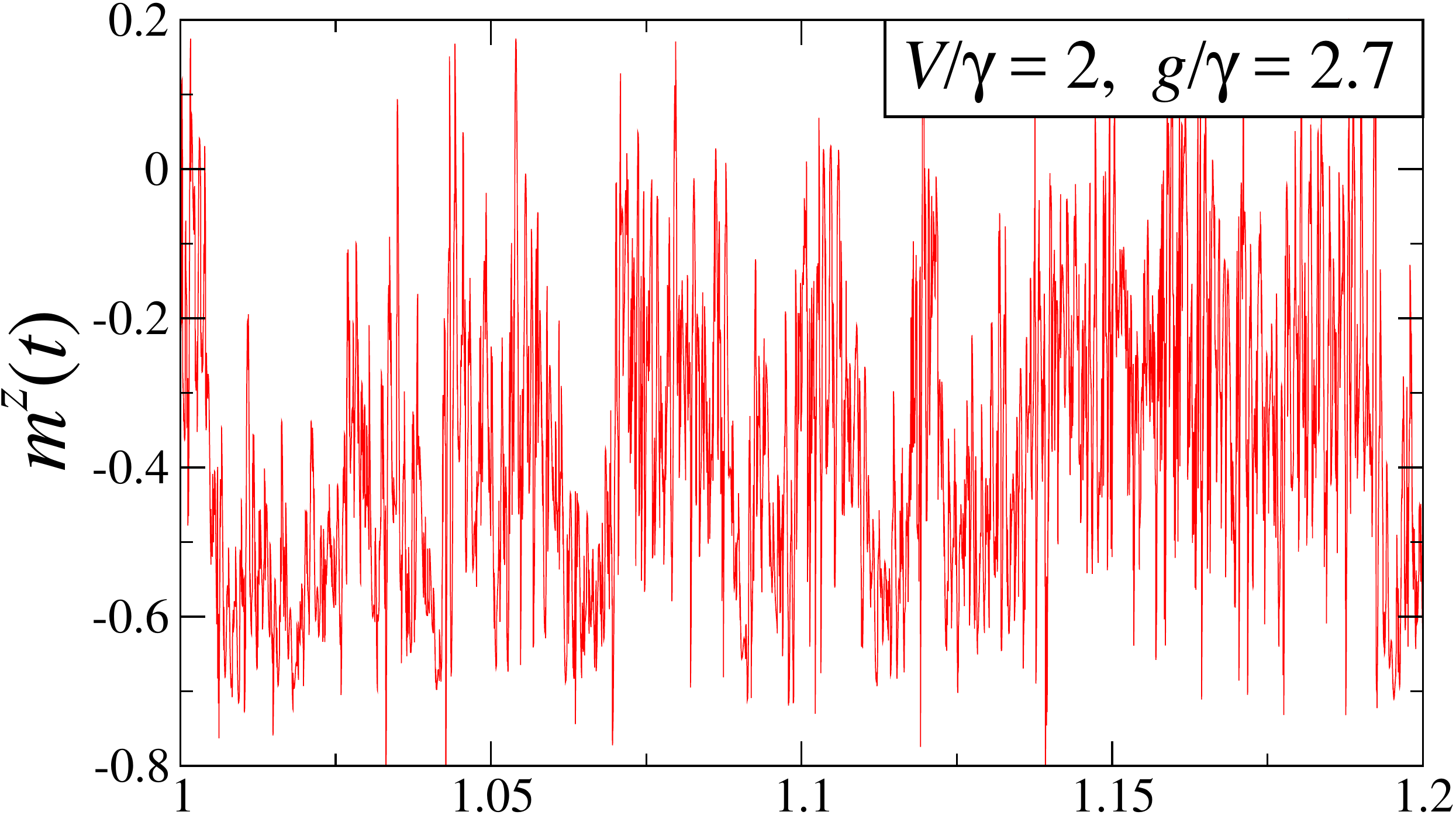}
  \includegraphics[width=0.95\columnwidth]{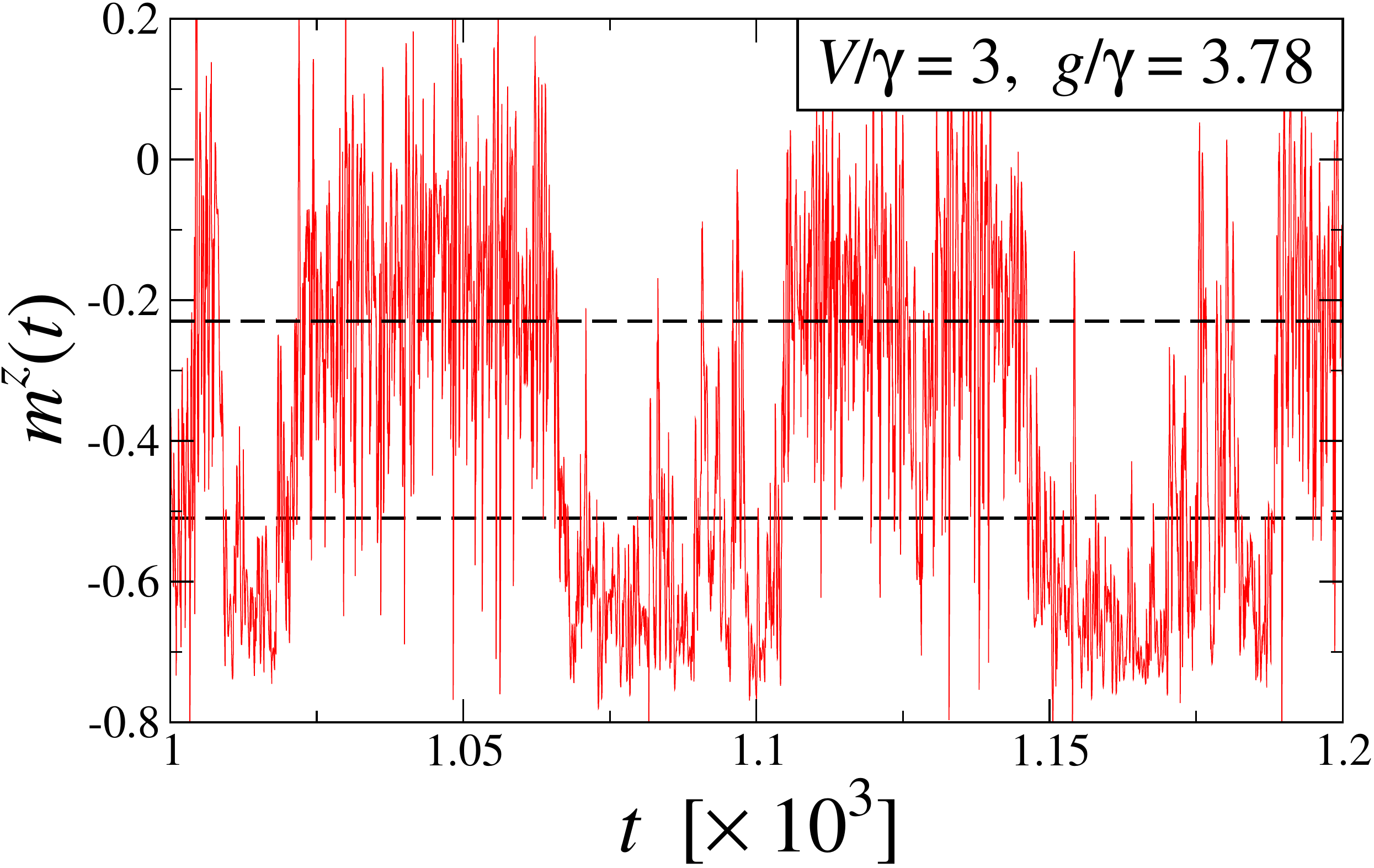}
  \caption{{\it Model I.} --- Time evolution of the magnetization $m^z(t)$ over
    a single quantum trajectory, evaluated in one of the central sites of the cluster.
    Here we employed a CMF approach with a cluster of size $\ell=4$.
    The upper panel is for $V/\gamma = 2$ and $g/\gamma = 2.7$ (non-critical region),
    while the lower panel is for $V/\gamma = 3$ and $g/\gamma = 3.78$
    (bistability region). Dashed black lines denote the two reached magnetization values,
    after averaging over many trajectories (see the upper right panel of Fig.~\ref{fig:Fig1}).}
  \label{fig:QTraj}
\end{figure}

We have analyzed the time evolution of a single quantum trajectory, looking
at the temporal behavior of the corresponding magnetization, close to the first-order transition.
In Fig.~\ref{fig:QTraj} we show two interesting situations, where bimodality can be observed or not.
The upper panel refers to a configuration where $V<V_c$, and $g$ corresponds
to the value in which $\partial m^z(g)/ \partial g$ is maximum (see Sec.~\ref{sec:modelI}).
In this case, during its evolution, the trajectory unravels a wide range of values of $m^z$, giving rise 
to the probability distribution shown in the lower left panel of Fig.~\ref{fig:Fig1}.
Here we do not observe any specific feature of the time-resolved measure, thus signaling
the absence of a critical behavior.
On the opposite, the lower panel corresponds to a situation where $V>V_c$, and $g$ lies
inside the (small) bistable region ($g_1 < g < g_2$) at finite cluster size.
In such case, it is evident that the trajectory may occasionally jump between the low-
and the high-polarized phase, thus giving rise to a bimodal distribution.
The typical switching time between the two metastable states is given by the inverse of the Liouvillian gap~\cite{Macieszczak_2016, Rose_2016}.

\section{Linear stability analysis for Model II}
\label{app:modelII}

An effective way to check the robustness of the CMF results for {\it Model II},
close to the continuous phase transition, is the linear stability analysis
over the CMF solution. Here we briefly sketch the procedure;
for further details, see Refs.~\cite{LeBoite_2013, Jin_2016}.
We recall that the CMF approach is based on a cluster-product ansatz for
the system's density matrix:
\begin{equation}
  \rho(t) = \bigotimes_{\cal C} \rho_{\cal C}(t) \, .
\end{equation}
The net computational advantage of CMF is that, within this approximation, the problem
can be reduced to solving an effective master equation of the type in Eq.~\eqref{eq:Master},
for the cluster density matrix $\rho_{\cal C}(t)$. The many-body Hamiltonian
$\hat H$ has to be replaced by an effective cluster Hamiltonian $\hat H_{\rm eff}$,
following the usual mean-field decoupling procedure.
For example, if we consider a square cluster of size $\ell \times \ell$,
the effective Hamiltonian $\hat H_{\rm eff}$ will be restricted to an interacting
model of $\ell^2$ sites.

\begin{figure}[!t]
  \centering
  \includegraphics[width=0.94\columnwidth]{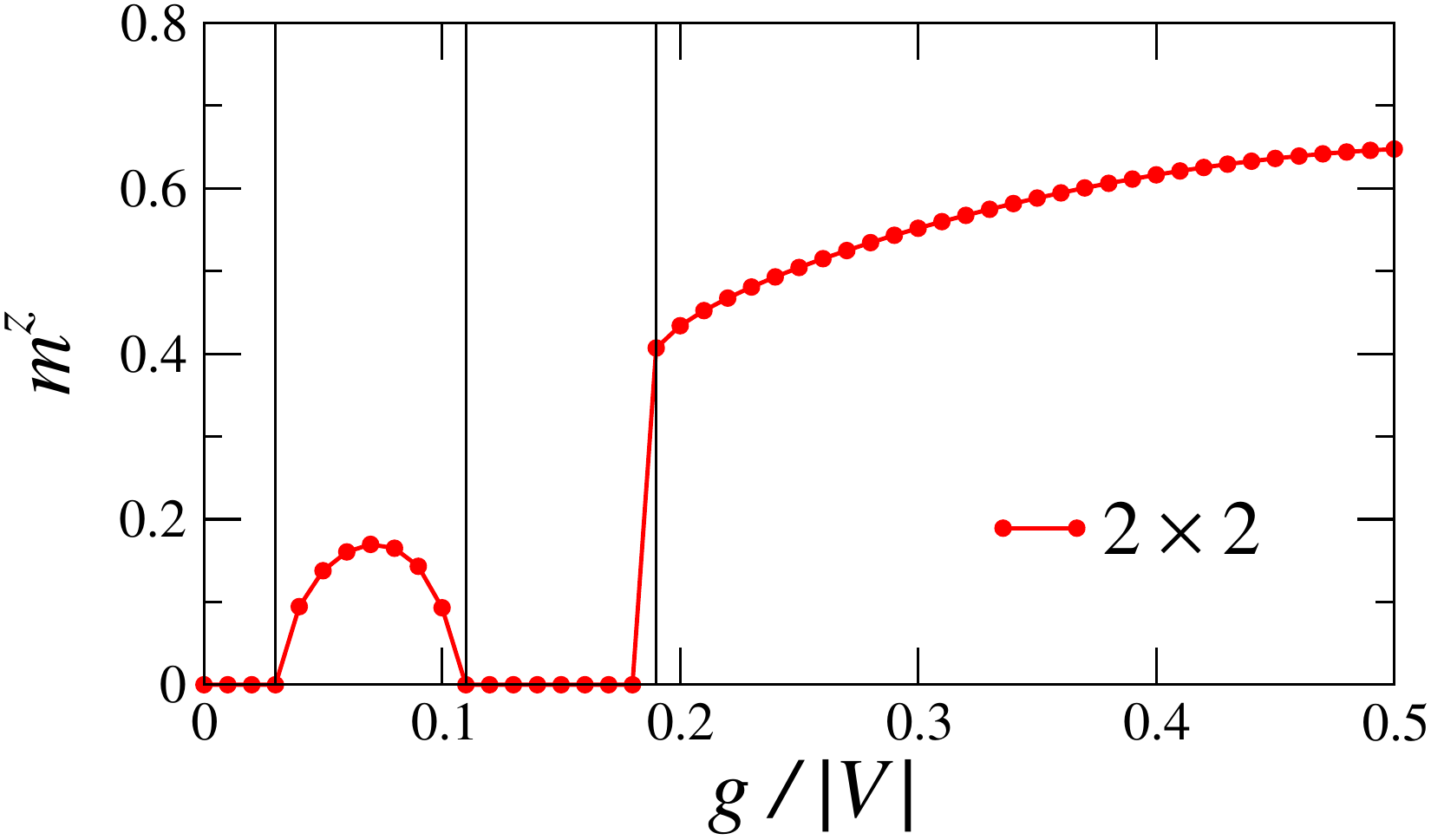} \hspace*{0.2cm}
  \includegraphics[width=0.99\columnwidth]{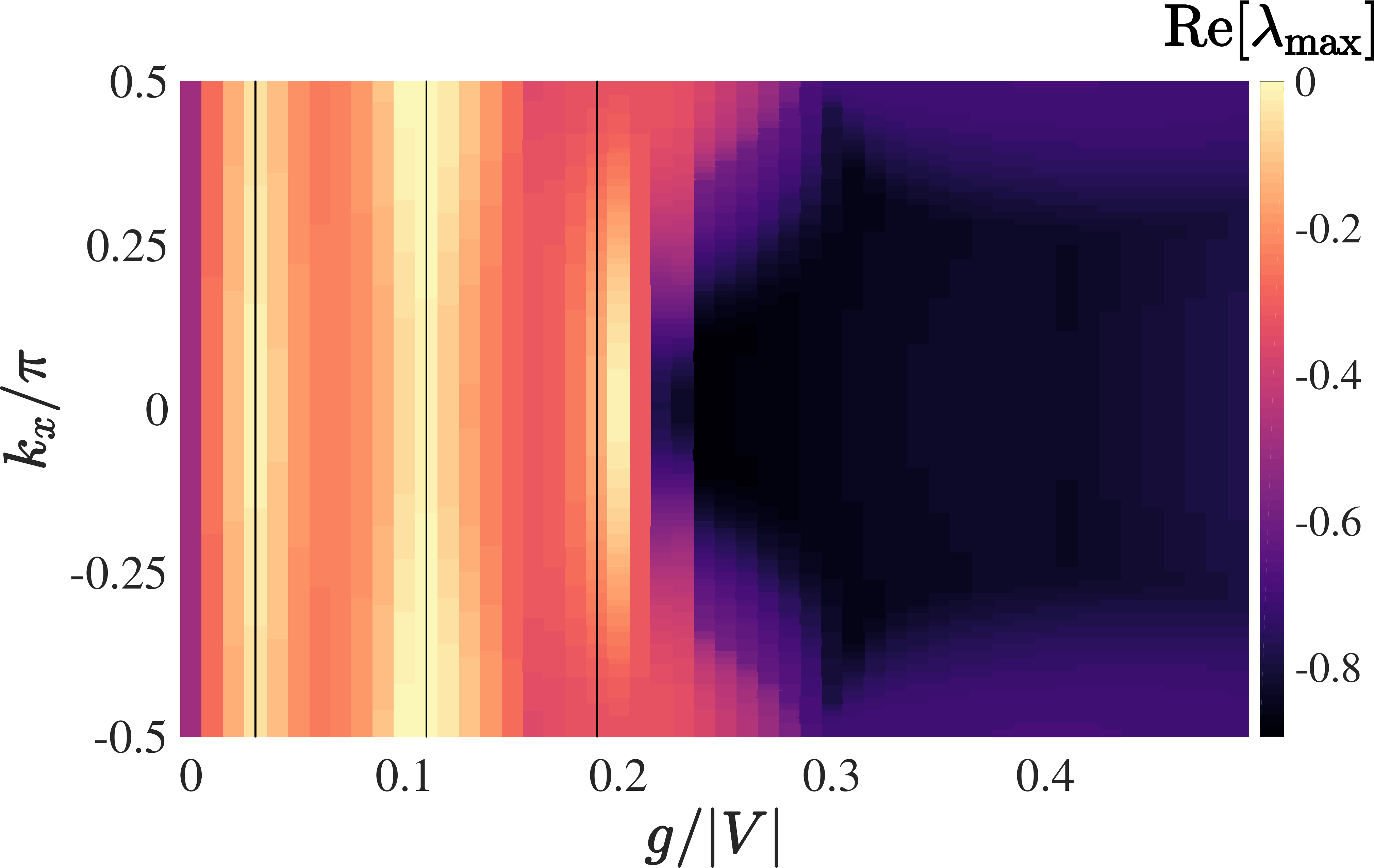}
  \caption{{\it Model II.} --- Upper panel: CMF analysis of the average magnetization
    $m^z$ as a function of $g/|V|$, and fixed $\gamma / |V| = 0.2$.
    Data are for a cluster of size $\ell=2$ (see red circles in the central panel of Fig.~\ref{fig:model2}).
    Bottom panel: linear stability analysis of the above results: real part of the most unstable 
    eigenvalue (negative means stable) as a function of $g/|V|$ and $k_x/\pi$, for $k_y = 0$.
    The vertical black lines are drawn in correspondence of the three relevant cuts 
    where the system may develop a putative phase transition.}
  \label{fig:StabAn}
\end{figure}

The linear stability analysis is performed on the factorized density matrix $\rho_{\cal C}(t)$.
To this purpose, it is useful to rewrite the equation of motion for the $j$-th
cluster in the superoperator formalism, according to
\begin{equation}
  \partial_t |\!|\rho_j \rangle\!\rangle = \hat {\mathbb{M}}_0 |\!|\rho_j \rangle\!\rangle + \sum_l
  \Big( \hat {\mathbb{E}}_l \cdot |\!|\rho_{j + {\bf e_l}} \rangle\!\rangle \Big) \: \hat {\mathbb{M}}_l |\!|\rho_j \rangle\!\rangle .
\end{equation}
Here $|\!|\rho_j\rangle\!\rangle$ is the vectorized form of the density matrix $\rho_j$ for the $j$-th cluster,
while $\hat{\mathbb{A}}_j$ denotes a given superoperator. In the above formula,
$\hat{\mathbb{M}}_0$ groups all the on-cluster terms, while $\hat{\mathbb{M}}_l$
is the on-cluster part of an off-cluster term and $\hat{\mathbb{E}}_l$
the corresponding off-cluster expectation.
The vector ${\bf e_l}$ is the direction to the neighboring cluster involved.

The linear stability analysis is performed by expanding the (small) fluctuations on top of the steady-state solution in terms of plane waves:
\begin{equation}
  |\!|\rho_j\rangle\!\rangle = |\!|\rho_0\rangle\!\rangle + \sum_{\bf k} e^{i {\bf k}\cdot{\bf r_j}} |\!|\delta \rho_{\bf k}\rangle\!\rangle ,
\end{equation}
in such a way that the resulting equation of motion for $|\!|\delta \rho_{\bf k}\rangle\!\rangle$
can be eventually cast into the form
\begin{equation}
  \partial_t |\!|\delta \rho_{\bf k} \rangle\!\rangle = \bigg[ \hat{\mathbb{M}}^{0,{\rm eff}} + \sum_{\bf e} e^{i{\bf k}\cdot{\bf e}}
    \: \hat{\mathbb{M}}^{1,{\bf e}} \bigg] |\!|\delta \rho_{\bf k}\rangle\!\rangle .
  \label{eq:superop}
\end{equation}
In the latter expression we have grouped together terms with the same vector ${\bf e}$,
as these all get the same ${\bf k}$-dependent factor.
Finally one can compute the eigenvalues of the effective superoperator contained in
the square brackets in Eq.~\eqref{eq:superop}, for each value of ${\bf k} = (k_x, k_y)$.
The eigenvalue with the largest positive real part represents the most unstable one.
Since the vectors ${\bf e_l}$ must be $\ell$ times
the elementary lattice vectors, the range of lattice momenta is restricted to the first
Brillouin zone of the superlattice, that is, $|k_j| < \pi / \ell$.

Numerical data containing the real part of the most unstable eigenvalue
for the effective superoperator of Eq.~\eqref{eq:superop} are shown in Fig.~\ref{fig:StabAn}.
There we display the outcome of the stability analysis for a CMF approach with $\ell=2$, following 
the same cut in the parameters space that has been analyzed in details in Sec.~\ref{sec:modelII}
($\gamma/|V| = 0.2$).
As is clearly visible from the contour plot at fixed $k_y=0$, in the region $g/|V| \lesssim 0.1$
the steady-state solution gets unstable for a wide range of $k_x$ in proximity of the boundaries of the tiny bump in the magnetization. 
In particular the instability at $|k_x|=|{\bf k}|=\pi/2$ suggests that the finite cluster size (which imposes this {\it artificial} periodicity) is responsible for such magnetic region, which has to be considered an artifact of the CMF ansatz.
Indeed, as the size of the cluster is increased, the value of the order parameter decreases
(see the central panel of Fig.~\ref{fig:model2}). 

On the other hand, the formation of a magnetized region for larger values of $g$ is robust.
Notice that, close to the value $g_c/|V| \sim 0.2$ corresponding to the critical point, our analysis predicts 
the onset of instability only for $k_x \simeq 0$. This witnesses the emergence of a homogeneous stationary solution
that spontaneously breaks the symmetry of the model, and gives rise to a genuine magnetic phase in which $|m^z|>0$.


\begin{thebibliography}{100}
\bibitem{footnote1}
  For clusters of size $\ell = 2$ we were able to diagonalize exactly the Liouvillian.
  For $\ell = 3$ the steady state has been obtained by simulating the time evolution of the full master equation
  with a fourth-order Runge-Kutta algorithm. For $\ell = 4$ we resorted to quantum trajectories.
  Unless specified, data presented in the paper have been obtained by averaging over $\gtrsim 500$ trajectories.
\end{thebibliography}
\end{document}